\documentclass[twocolumn,showpacs,showkeys,preprintnumbers,amsmath,amssymb]{revtex4}

\usepackage{graphicx}
\usepackage{dcolumn}
\usepackage{bm}

\input epsf

\begin{document}


\title{Magnetic Phase Transition at 88 K in Na$_{0.5}$CoO$_2$\\
       revealed by $^{23}$Na-NMR investigations}

\author{B. Pedrini, J. L. Gavilano, S. Weyeneth, E. Felder, J. Hinderer, M. Weller, H. R. Ott,
S. M. Kazakov, J. Karpinski}
\affiliation{
Laboratorium f\"{u}r Festk\"{o}rperphysik,\\
ETH-H\"{o}nggerberg, CH-8093~Z\"{u}rich, Switzerland
}

\date{\today}

\begin{abstract}
Na$_{0.5}$CoO$_{2}$ exhibits a metal-insulator transition at $T_{\mathrm{MI}}=53$ K upon cooling. The nature of another transition at $T_{\mathrm{X}}=88$ K has not been fully clarified yet.
We report the results of measurements of the electrical conductivity $\sigma$, the magnetic susceptibility $\chi$ and $^{23}$Na NMR on a powder sample of Na$_{0.5}$CoO$_{2}$, including the mapping of NMR spectra, as well as probing the spin-lattice relaxation rate $T_1^{-1}(T)$ and the spin-spin relaxation rate $T_2^{-1}(T)$, in the temperature range between 30 K and 305 K.
The NMR data reflect the transition at $T_{\mathrm{X}}$ very well but provide less evidence for the metal-insulator transition at $T_{\mathrm{MI}}$.
The temperature evolution of the shape of the spectra implies the formation of a staggered internal field below $T_{\mathrm{X}}$, not accompained by a rearrangement of the electric charge distribution.
Our results thus indicate that in Na$_{0.5}$CoO$_{2}$, an unusual type of magnetic ordering in the metallic phase precedes the onset of charge ordering, which finally induces an insulating ground state.
\end{abstract}


\maketitle

\section{\label{SecIntroduction}Introduction}

Compounds of the series Na$_{x}$CoO$_{2}$ ($0.25<x<1$) adopt a crystal structure which consists of alternating layers of Na$^+$ ions and layers of edge-sharing CoO$_6$ octahedra, with a triangular arrangement of the Co sites \cite{Huang_2004_PRB}.
They represent a physical realization of magnetic systems with planar triangular symmetry, in which metallicity is achieved by controlled carrier injection. 
The simple metallic phase ($x<0.5$) is separated from the local moment metallic phase ($x>0.5$) by the insulating phase occurring at $x=0.5$ \cite{Foo_2004_PRL}.
A variety of other interesting phenomena is observed at different values $x$ of Na-concentration upon varying the temperature.
The material with composition $x\approx0.7$ exhibits a large thermoelectric power \cite{Terasaki_1997_PRB}, and indications of a charge instability in the temperature range between 230 and 290 K were observed \cite{Gavilano_2004_PRB}.
A spin-density wave formation at temperatures below 22 K was detected for $x=0.75$ \cite{Sugiyama_2003_PRB}.
Superconductivity is induced at $x=0.35$, if the CoO$_2$-layers are intercalated with water \cite{Takada_2003_Nature}.

The richness of the phase diagram of Na$_{x}$CoO$_{2}$ is argued to be the result of a subtle interplay between electronic and structural degrees of freedom.
Neutron scattering measurements \cite{Huang_2004_PRB} revealed a critical dependence of both the shape of the CoO$_6$ octahedra and the interlayer distance on the Na concentration $x$ and on the distribution of the Na$^+$ ions in the intervening layers, where two possible sites, Na1 and Na2, are available.
The presence of ordered Na$^+$ ions and Na$^+$ vacancy superlattices beyond the simple average hexagonal structure was confirmed by means of electron diffraction measurements \cite{Zandbergen_2004_PRB}.

It is a peculiarity of the Na$_{x}$CoO$_{2}$-series to exhibit an insulating ground state in a very narrow range of values $x$ centered at $x=0.5$.
The metal-insulator transition in Na$_{0.5}$CoO$_{2}$, occurring at $T_{\mathrm{MI}}=53$ K \cite{Huang_2004_JPCM,Foo_2004_PRL} upon cooling, is reflected by anomalies in the temperature dependences of the resistivity $\rho(T)$, the magnetic susceptibility $\chi(T)$, and the specific heat. The insulating character of the low-temperature phase was also confirmed by infrared reflectivity measurements \cite{Wang_2004_PRL}.
Unusual features were observed for the thermal conductivity, the thermopower and the Hall resistivity $R_{\mathrm{H}}(T)$ \cite{Foo_2004_PRL}.
In addition, Na$_{0.5}$CoO$_{2}$ shows a pronounced Na superstructure \cite{Huang_2004_JPCM,Zandbergen_2004_PRB}, where the Na$^+$ ions order in zig-zag chains \cite{Huang_2004_PRB,Zhang_2005_PRB} with an alternating occupation of Na1 and Na2 sites.
Besides the metal-insulator transition, another transition at $T_{\mathrm{X}}=88$ K is clearly reflected by an abrupt reduction of $\chi(T)$ upon cooling. This transition is also characterized by a change of sign of $R_{\mathrm{H}}(T)$ \cite{Foo_2004_PRL}, which rapidly decreases to large negative values below $T_{\mathrm{X}}$ with decreasing temperature.
In Ref.\cite{Huang_2004_JPCM} the transition was argued to be of structural type.
Subsequent $\mu$SR experiments have indicated an antiferromagnetically ordered ground state, but different ordering temperatures of 53 K \cite{Uemura_2004_cond-mat} and 85 K \cite{Mendels_2005_PRL} were reported.
Very recent $^{59}$Co NMR and neutron scattering studies \cite{Yokoi_2005_cond-mat} brought evidence of antiferromagnetic order below 87 K. The order was claimed to involve two types of localized Co magnetic moments of 0.34 $\mu_B$ and 0.11 $\mu_B$, respectively.

The variety of experimentally observed features of Na$_{x}$CoO$_{2}$ triggered a number of intense theoretical investigations.
The electronic structure of these compounds was calculated in the local (spin) density approximation (LSDA). 
The inclusion of a Hubbard repulsion parameter $U$ in the calculation yields the two magnetically different regimes for $x<0.5$ and $x>0.5$ \cite{Lee_2004_PRB} that are observed experimentally.
The calculation also explains \cite{Zhang_2004_PRL} the absence of pockets of the Fermi surface, as was registered in angle-resolved photoemission (ARPES) experiments for $x=0.3$ \cite{Hasan_2005_cond-mat}, $x=0.6$ \cite{Yang_2004_PRL} and $x=0.7$ \cite{Hasan_2004_PRL}.
If the value of $U$ is chosen large enough, an insulating ground state can be obtained for Na$_{0.5}$CoO$_{2}$ \cite{Li_2005_PRB}. The authors of Ref. \cite{Lee_2005_PRL} describe a charge disproportionation accompanied by magnetic moment localization $2\mathrm{Co}^{3.5+}\rightarrow \mathrm{Co}^{3+}+\mathrm{Co}^{4+}$, occuring at a critical value $U=3.4$ eV.

In a different approach, a multiorbital Hubbard model with strong Coulomb repulsion was considered. The calculations predict both the Fermi surface pocket cancellation and a band narrowing \cite{Zhou_2005_PRL}, again in good agreement with the results of ARPES measurements \cite{Yang_2005_cond-mat} at $x\neq0.5$.
With a similar approach, the authors of Ref.\cite{Indergand_2005_PRB} focus on charge and spin instabilities, deriving a variety of possible metallic states with a spontaneous breaking of the original hexagonal symmetry; in particular, the zig-zag arrangement of Na$^+$ ions observed in Na$_{0.5}$CoO$_{2}$ is found to be energetically favourable and it is argued that this superstructure leads to a quasi one-dimensional band structure, with the possible formation of a spin-density wave.

The aim of our work was to elucidate the nature of both the phase transition at $T_{\mathrm{X}}$ and the phase between $T_{\mathrm{MI}}$ and $T_{\mathrm{X}}$ in Na$_{0.5}$CoO$_{2}$.
It is predominantly our $^{23}$Na NMR data that give clear evidence for a transition from a paramagnetic to an antiferromagnetically ordered state at $T_{\mathrm{X}}=88$ K upon cooling.
This transition occurs in a low-conductivity metallic phase and does not result in the opening of a gap in the spectrum of electronic excitations.
This observation is, to some extent, surprising because a magnetic transition is rather expected to occur in the low-temperature insulating phase, or simultaneously with the metal-insulator transition at $T_{\mathrm{MI}}$.
To our knowledge, no theoretical model that would account for the described phenomenon has yet been proposed.

The paper is organised as follows.
The preparation of the Na$_{0.5}$CoO$_{2}$ powder sample is described in Sec. \ref{SecSample}.
In Sec. \ref{SecSigma} and \ref{SecChi}, we report the results of the measurements of the electrical conductivity $\sigma(T)$ and of the magnetic susceptibility $\chi(T)$.
Section \ref{SecNaNMR} is devoted to the $^{23}$Na NMR experiments.
We present the temperature variations of the spectra as well as of the spin-lattice and spin-spin relaxation rates $T_1^{-1}$ and $T_2^{-1}$.
In Sec. \ref{SecResults}, we discuss the experimental results.
The emphasis is put on the magnetic properties of the phases at temperatures below $T_{\mathrm{X}}$, but also the unusual properties of the metallic regime above $T_{\mathrm{X}}$ are considered to some extent.

\section{\label{SecSample}Sample}

The Na$_{0.5}$CoO$_{2}$ powder sample was prepared by a procedure that is described in Ref.\cite{Huang_2004_JPCM}.
First, a powder of nominal composition Na$_{0.75}$CoO$_{2}$ was prepared by heating a stoichiometric mixture of Co$_3$O$_4$ (Aldrich, 99.995$\%$) and Na$_2$CO$_3$ (Aldrich, 99.995$\%$) in air at 850$^{\circ}$C for 15 hours. The remelting procedure was repeated twice to ensure the proper cation concentration.
For the synthesis of Na$_{0.5}$CoO$_{2}$, 0.5 g of Na$_{0.75}$CoO$_{2}$ were stirred in 120 ml of a solution of I$_2$ in acetonitrile for 4 days.
The quantity of dissolved I$_2$ was chosen to be 10 times the amount that would in principle be necessary to remove all the Na from Na$_{0.75}$CoO$_{2}$.
The resulting material was cleaned with acetonitrile and dried in flowing argon. X-ray powder diffraction patterns showed that the resulting material is of single phase with lattice parameters of $a$=2.8134(7) and $c$=11.129(4) \AA.
These parameters yield an estimated Na content of $x$=0.48 \cite{Foo_2004_PRL}.

\section{\label{SecSigma}Electrical conductivity}

\begin{figure}[t]
 \begin{center}
  \leavevmode
  \epsfxsize=1.0\columnwidth \epsfbox{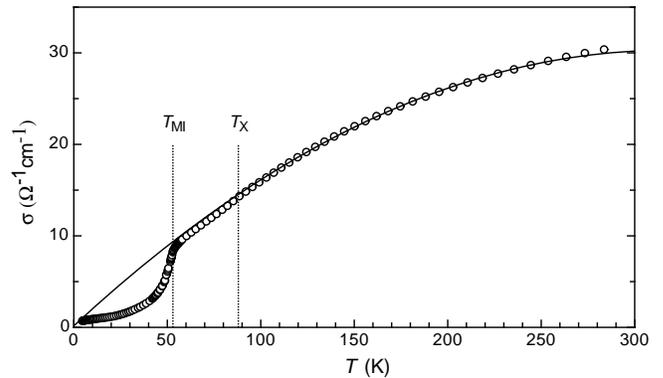}
\caption{
Electrical conductivity $\sigma$ as a function of temperature $T$.
The solid line represents the results of best fits with the functions $\sigma(T)=c_2T^{2}+c_1T+c_0$. Data in the temperature range 100-260 K were considered.
}
\label{FigSigma}
\end{center}
\end{figure}

As a first characterization of our sample we measured the temperature dependence of the electrical resistivity.
The resistance $R$ of a piece of pressed powder, whose length and cross section were $d\approx 6$ mm and $A\approx15$ mm$^{2}$ respectively, was measured using an a.c. four-terminal technique.

The conductivity $\sigma=(RA/d)^{-1}$ is shown, as a function of temperature $T$, in Fig. \ref{FigSigma}.
The values of $\sigma$ are, on the average, smaller by a factor of 30 than $\sigma_{\parallel}$ measured along the CoO$_2$-planes of a single crystal and reported in Ref.\cite{Huang_2004_JPCM}.
Nevertheless, the temperature dependence of both quantities is similar.
While electron scattering at the boundaries of the powder grains is expected to occur in our sample, we do not identify it as the main cause of the discrepancy between $\sigma$ and $\sigma_{\parallel}$.
The large difference and the similar temperature dependence instead suggest a significant anisotropy of the conductivity in Na$_{0.5}$CoO$_{2}$, not unexpected in a layered compound (see, e.g., La$_{0.5}$Sr$_{1.5}$MnO$_4$, where the anisotropy in $\sigma$ is of the order of 1000 \cite{Moritomo_1995_PRB}).

Upon cooling, the gradual decrease of $\sigma$  at high temperatures (a factor of 5 between 290 K and 100 K) is intecepted by an abrupt downturn at $T_{\mathrm{MI}}=53$ K.
This reflects the metal-insulator transition.

The high-temperature phase is not properly metallic.
The continuous decrease of $\sigma(T)$ with decreasing temperature is certainly not typical for a metal, but often observed in oxide materials \cite{Imada_1998_RMP}.
Further insight into the transport properties can be gained by estimating the mean free path $l$ of the charge carriers. From the Drude law
\begin{equation}
  \label{EqDrude}
  \sigma=\frac{e^{2}n_{\mathrm{e}}\tau}{m^{\ast}}=\frac{e^{2}n_{\mathrm{e}}l}{\hbar k_F}
  \quad,
\end{equation}
with $n_{\mathrm{e}}$ as the charge carrier density, $\tau$ the average scattering time, $m^{\ast}$ the effective mass of the charge carriers and $k_F=m^{\ast}l/\hbar\tau$ the Fermi wave vector.
We assume one conduction electron per Na atom, resulting in $n_{\mathrm{e}}=1.31\times10^{+22}\mathrm{cm}^{-3}$.
For the conductivity $\sigma$ we insert the high-temperature saturation value $\bar{\sigma}$, which is calculated as follows.
The temperature dependence of the conductivity $\sigma(T)$ for temperatures in the range between 100 and 260 K can be well approximated by (see Fig. \ref{FigSigma})
\begin{equation}
  \label{EqQuadraticSigma}
  \sigma(T)=c_2 T^{2}+c_1 T+c_0
\end{equation}
with $c_2=-2.98\times10^{-4}\;\mathrm{K}^{-2}(\Omega\mathrm{cm})^{-1}$, $c_1=0.19\;\mathrm{K}^{-1}(\Omega\mathrm{cm})^{-1}$, and $c_0\approx0$.
Accordingly, the maximum of $\sigma$ is reached at $\bar{T}=318$ K and we take
\begin{equation}
  \label{EqSigmaSaturated}
  \bar{\sigma}=\sigma(\bar{T})\approx30\;(\Omega\mathrm{cm})^{-1}
\end{equation}
as an estimate of the saturation value at high temperatures.

We first assume a three-dimensional (3D) free electron gas with
\begin{equation}
  k_F = (3\pi^{2} n_{\mathrm{e}})^{\frac{1}{3}} \quad(\mathrm{3D})
  \quad.
\end{equation}
Inserting (\ref{EqSigmaSaturated}) into (\ref{EqDrude}) gives a mean free path of
\begin{equation}
  \label{EqkMFP3D}
  l = 6.9\times10^{-10}\mathrm{cm}\quad(\mathrm{3D})
  \quad,
\end{equation}
which is far less than the interatomic distance.

For a more realistic estimate of the mean free path $l$ we take into account the layer structure of the material and we thus assume a gas of electrons that is constrained to two-dimensional (2D) planes, i.e.,
\begin{equation}
    k_F  = (2\pi n_{\mathrm{e,2D}})^{\frac{1}{2}}    \quad(\mathrm{2D})
  \quad,
\end{equation}
where $n_{\mathrm{e,2D}}=n_{\mathrm{e}} c/2$ is the 2D electron density and $c/2$ the distance between two such planes ($c$ is the lattice constant along the $c$-axis, see Sec. \ref{SecSample}).
Thus, in Eq. (\ref{EqDrude}) we replace $\sigma$ with the estimated in-plane saturation value,
\begin{equation}
  \label{EqSigmaSaturated2D}
  \bar{\sigma}_{\parallel}\approx30\bar{\sigma}=900\;(\Omega\mathrm{cm})^{-1}
  \quad,
\end{equation}
taking into account the single-crystal data.
This yields a mean free path in the $ab$-plane of
\begin{equation}
  \label{EqMeanFreePath2D}
  l\approx1.9\times10^{-8}\mathrm{cm}^{-1} \quad(\mathrm{2D})
  \quad,
\end{equation}
and 
\begin{equation}
  \label{EqkFl2D}
  k_F l \approx 1.3 \quad(\mathrm{2D})
  \quad.
\end{equation}
Both these values suggest that a description of the electronic transport in the metallic phase in terms of simple scenarios is not tenable.
Further discussions on this point are postponed to Sec. \ref{SecResults}.

\section{\label{SecChi}Magnetic susceptibility}

\begin{figure}
 \begin{center}
  \leavevmode
  \epsfxsize=1.0\columnwidth \epsfbox{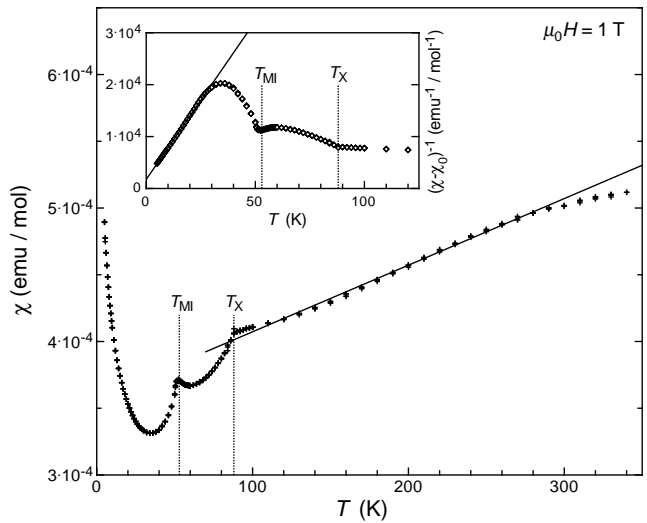}
\caption{
Magnetic susceptibility $\chi$ as a function of temperature $T$, measured in a fixed magnetic field $\mu_0H=1$ T. The solid line is the best linear fit for 100 K $<T<$ 260 K.
Inset: inverse magnetic susceptibility $(\chi-\chi_0)^{-1}$ as a function of $T$.
The solid line is the best linear fit for 5 K $<T<$ 20 K.
}
\label{FigChi}
\end{center}
\end{figure}

As an additional characterization of our sample, we measured the temperature dependence of the magnetic susceptibility $\chi(T)$.
The magnetization of 50 mg of Na$_{0.5}$CoO$_{2}$, again in the form of a powder sample (number of mols $n=4.88\times10^{-4}$), was measured in different magnetic fields up to $\mu_0H=5$ T and in the temperature range between 5 and 340 K.
At fixed temperatures the magnetic susceptibility $\chi=M/n\mu_0H$ is found to be field independent.
Figure \ref{FigChi} shows $\chi(T)$ for $\mu_0H=1$ T. Our result is in good agreement with previously published single-crystal data \cite{Huang_2004_JPCM,Yokoi_2005_cond-mat}.

Two anomalies at $T_{\mathrm{MI}}=53$ K and $T_{\mathrm{X}}=88$ K are revealed by kinks, both followed by abrupt decreases of $\chi$ upon cooling.

Above $T_{\mathrm{X}}$, $\chi(T)$ decreases continuously with decreasing temperature.
In the range from 260 to 100 K, the susceptibility depends approximately linearly on the temperature (see Fig. \ref{FigChi}).
Above 260 K, the experimentally measured susceptibility tends to saturate to an estimated value of
\begin{equation}
  \label{EqChiSaturated}
  \bar{\chi}=5.2\times10^{-4}\;\mathrm{emu/mol}
  \quad.
\end{equation}
Assuming that the susceptibility above 300 K is due to pure Pauli paramagnetism, $\bar{\chi}$ yields the total density of states at the Fermi energy
\begin{equation}
  \label{EqDOSChi}
  D(E_F)=\frac{1}{\mu_B^{2}V_{\mathrm{mol}}}\bar{\chi}
        =2.63\times10^{+35}\;\mathrm{erg}^{-1}\mathrm{cm}^{-3}
  \quad.
\end{equation}
Here, $V_{\mathrm{mol}}=22.97$ cm$^{3}$ is the molar volume of Na$_{0.5}$CoO$_{2}$.
The expected values of $D(E_F)$ for a free 3D electron gas and for a gas of electrons constrained to 2D planes, respectively, are
\begin{eqnarray}
  \label{EqDOS3D}
  D(E_F) & = &  2\frac{3n_{\mathrm{e}}^{\frac{1}{3}}m^{\ast}}{(3\pi^{2})^{\frac{2}{3}}\hbar^{2}}
  \quad(\mathrm{3D})
\end{eqnarray}
and
\begin{eqnarray}
  \label{EqDOS2D}
  D(E_F) & = &  2\frac{m^{\ast}}{(2\pi)\hbar^{2}\frac{c}{2}}
  \quad  (\mathrm{2D}),
\end{eqnarray}
where $c$ is again the value of the lattice constant along the $c$-axis (see Sec. \ref{SecSample}), and $m^{\ast}$ is the effective mass of the conduction electrons.
The comparison of $D(E_F)$ given in Eq. (\ref{EqDOSChi}) with (\ref{EqDOS3D}) and (\ref{EqDOS2D}) yields an effective mass
\begin{equation}
  \label{EqEffMass3D}
  m^{\ast}\approx22\; m_{\mathrm{e}} \quad (\mathrm{3D}),
\end{equation}
or, for the more realistic 2D situation,
\begin{equation}
  \label{EqEffMass2D}
  m^{\ast}\approx56\; m_{\mathrm{e}} \quad (\mathrm{2D}),
\end{equation}
where $m_{\mathrm{e}}$ is the free electron mass.
Further comments concerning these values are presented in Sec. \ref{SecResults}.

We now consider the possibility that the anomaly in $\chi(T)$ at $T_{\mathrm{X}}$ is due to antiferromagnetic ordering of localized magnetic moments upon cooling the sample.
In a very simple scenario, we suppose that the valences are Na$^{1+}$ and O$^{2-}$, and that there is one electron per Na atom in the conduction band. 
This leads to the $3d^5$ conficuration Co$^{4+}$. 
If we also assume that only the lowest $t_{2g}$ Co orbitals are occupied \cite{Zhou_2005_cond-mat}, we find that the total spin is $S=1/2$ per Co ion.
If the orbital component is quenched, the corresponding effective magnetic moment per formula unit is
\begin{equation}
  \label{EqMuEffLocMom}
  \mu_{\mathrm{eff},S=1/2}=g\sqrt{S(S+1)}=1.73
  \quad,
\end{equation}
where $g=2$.
The Curie-Weiss temperature dependence of the susceptibility is 
\begin{equation}
  \chi_{S=1/2}(T)=(\mu_{\mathrm{eff},S=1/2})^2\;\frac{N\mu_B^2}{3k_B}\frac{1}{T}
  \quad.
\end{equation} 
At $T$= 300 K, $\chi_{S=1/2}=1.25\times10^{-3}\;\mathrm{emu/mol}$, and at the transition temperature $T_{\mathrm{X}}$ $\chi_{S=1/2}=4.26\times10^{-3}\;\mathrm{emu/mol}$, respectively.
These values are much larger than the measured values of $\chi$, thus indicating that no significant moments reside on the Co sites.
The susceptibility $\chi(T)$ for $T\geq T_{\mathrm{MI}}$ is thus attributed, as claimed above, to the Pauli paramagnetism of the itinerant charge carriers.
In previous work \cite{Yokoi_2005_cond-mat}, the reduction of $\chi$ upon cooling below room temperature was ascribed to a pseudogap feature in the electronic excitation spectrum.

Below 20 K, $\chi(T)$ increases with decreasing temperature.
Forcing $\chi(T)$ for 5 K $\leq T\leq$ 10 K to fit a Curie-Weiss function $ \chi_0+C(T-\Theta)^{-1}$ results in $\chi_0=2.82\times10^{-4}$ emu/mol, significantly less than above room temperature.
The inset of Fig. \ref{FigChi} shows $(\chi-\chi_0)^{-1}$ as a function of $T$, emphasizing that
\begin{equation}
  \label{EqCurieWeissInv}
  (\chi-\chi_0)^{-1}\approx C^{-1}(T-\Theta)
\end{equation}
up to 30 K, with $C=(1.7\pm0.1)\times10^{-3}$ (emu mol)$^{-1}$ K and $\Theta=2.5\pm0.4$ K.
The effective magnetic moment per formula unit is thus (in units of $\mu_B$)
\begin{equation}
  \label{EqMuEff}
  \mu_{\mathrm{eff}}=\sqrt{\frac{3k_B}{N\mu_B^{2}}C}=0.11\pm0.005
  \quad.
\end{equation}
The comparison of the latter value with the value calculated in Eq. (\ref{EqMuEffLocMom}) leads to the conclusion that at most 7$\%$ of the Co atoms carry  magnetic moments, which are still acting freely at low temperatures.

\section{\label{SecNaNMR}$^{23}\mathrm{Na}$ NMR}

The $^{23}\mathrm{Na}$ NMR experiments were made with two samples of pressed powder with masses 45 and 325 mg, respectively.
No significant differences between the results of the measurements were observed.

The values of the magnetic fields $\mu_0H$ in which the experiments were performed were determined from the measured resonance frequency of deuterated water D$_2$O.
The $^{23}\mathrm{Na}$ nuclear reference frequency $f_0$ was calculated using the standard gyromagnetic ratio $\gamma_{\mathrm{Na}}$, i.e., $f_0=\gamma_{\mathrm{Na}}\mu_0H$.

\subsection{Spectra}

$^{23}$Na NMR spectra were obtained by monitoring the integrated spin-echo intensity as a function of the irradiation frequency in a fixed magnetic field $H$. The echo was generated with a two-pulse $\pi/2$-delay-$\pi$ spin-echo sequence, irradiating a frequency window of about 25 kHz.

\begin{figure}
 \begin{center}
  \leavevmode
  \epsfxsize=1.0\columnwidth \epsfbox{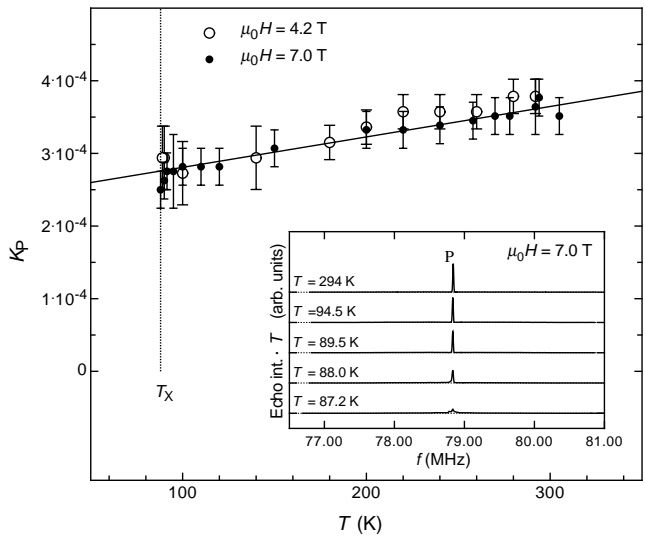}
\caption{
Relative frequency shift $K_{\mathrm{P}}$ as a function of temperature $T$ of the $^{23}$Na NMR line P ($T\geq88$ K), in magnetic fields $\mu_0H=4.2$ T and $\mu_0H=7.0$ T.
Inset: $^{23}$Na NMR spectra measured in a fixed magnetic field of $\mu_0H=7.0$ T, for $T\geq87.2$ K.
The integrated echo-intensity, multiplied with the temperature $T$, is plotted as a function of the irradiation frequency $f$.
}
\label{FigKTParam}
\end{center}
\end{figure}

In the inset of Fig. \ref{FigKTParam}, we show examples of the spectra taken in a magnetic field $\mu_0H=7.0$ T at temperatures above 87.2 K.
Above $T_{\mathrm{X}}=88$ K, we observe a single, narrow NMR line, labeled P, peaking at a frequency $f_{\mathrm{P}}$.
The relative frequency shift $K_{\mathrm{P}}=(f_{\mathrm{P}}-f_0)/f_0$ is represented in Fig. \ref{FigKTParam} as a function of temperature $T$.
$K_{\mathrm{P}}$ is field independent and, above $T_{\mathrm{X}}$, varies approximately linearly with $T$. It is related to the susceptibility $\chi$ by
\begin{equation}
  \label{EqKChi}
  K_{\mathrm{P}}\approx\frac{1}{N\mu_B}A_{\mathrm{P}}\;\chi+K_0
  \quad.
\end{equation}
The parameter $K_0\approx0$ and the hyperfine coupling constant is  $A_{\mathrm{P}}=0.43\pm0.03$ T per $\mu_B$ magnetization of a formula unit of Na$_{0.5}$CoO$_{2}$.
The full width at half maximum (FWHM) of the NMR line P, $\Delta f_{\mathrm{P}}$, slowly increases upon cooling from the room temperature value, $\Delta f_{\mathrm{P}}\approx12$ kHz, to $\Delta f_{\mathrm{P}}\approx16$ kHz at $T_{\mathrm{X}}$.
\begin{figure}
 \begin{center}
  \leavevmode
  \epsfxsize=1.0\columnwidth \epsfbox{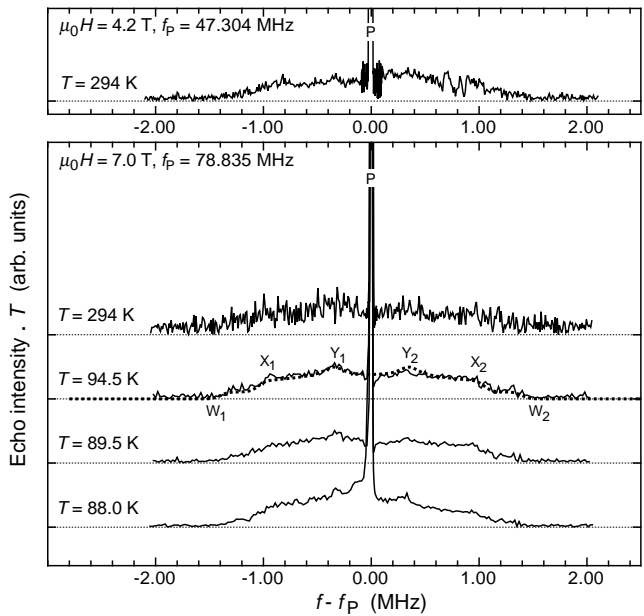}
\caption{
$^{23}$Na NMR spectra measured in  fixed magnetic fields $\mu_0H=4.2$ T (upper plot) and $\mu_0H=7.0$ T (lower plot), for $T\geq88.0$ K.
The integrated echo-intensity, multiplied with the temperature $T$, is plotted as a function of $f-f_{\mathrm{P}}$, with $f$ the irradiation frequency and $f_{\mathrm{P}}$ the frequency of the central line P.
The dotted line on the spectrum at $T=94.5$ K represents the calculated shape of the wings due to first-order quadrupolar effects.
The signifiance of the different letters is explained in the text.
}
\label{FigQuadParam}
\end{center}
\end{figure}

The spectra recorded at temperatures $T\geq T_{\mathrm{X}}$ exhibit wings in the range labeled W$_1$-W$_2$ with features labeled X$_1$, $Y_1$ and X$_2$, $Y_2$.
Corresponding spectra are shown in Fig. \ref{FigQuadParam}.
The shape and the width $\Delta f_{\mathrm{W}}=f_{\mathrm{W}_2}-f_{\mathrm{W}_1}$ of the wings are the same for $\mu_0H=4.2$ T and $\mu_0H=7.0$ T, suggesting that they represent the quadrupolar wings of peak P.
The ratio between the intensities of the central line P and the wings is $0.62\pm0.7$, close to the theoretical value of $2/3$.
Our interpretation is further supported by the results of comparisons of the measured signals with the signal calculated by assuming a random orientation of the crystalline grains and first order quadrupolar shifts of the transitions $\pm\frac{3}{2}\leftrightarrow \pm\frac{1}{2}$ for spin $I=3/2$ nuclei according to \cite{Abragam}
\begin{equation}
  \Delta f_Q^{(1)}=\pm\frac{f_Q}{2}
                   \left[3\cos^2\theta-1-\eta\sin^2\theta\cos2\phi\right]
  \quad.
\end{equation}
Here, $\theta$ and $\phi$ are the spherical angles describing the orientation of the principal axes of the electric field gradient tensor with respect to the direction of the applied magnetic field.
The dotted line in Fig. \ref{FigQuadParam} for $T=94.5$ K represents the calculated signal without the contribution of the central transition $+\frac{1}{2}\leftrightarrow -\frac{1}{2}$ (the peak labeled P).
The best coincidence is achieved with a quadrupolar frequency $f_Q=1/2\Delta f_{\mathrm{W}}= 1.4$ MHz and an asymmetry parameter $\eta=0.5$.
No substantial variations in the shape and the width of the wings were found in the temperature range between 295 and 88 K, suggesting only very small rearrangements of electrical charge above $T_{\mathrm{X}}$.

\begin{figure}
 \begin{center}
  \leavevmode
  \epsfxsize=1.0\columnwidth \epsfbox{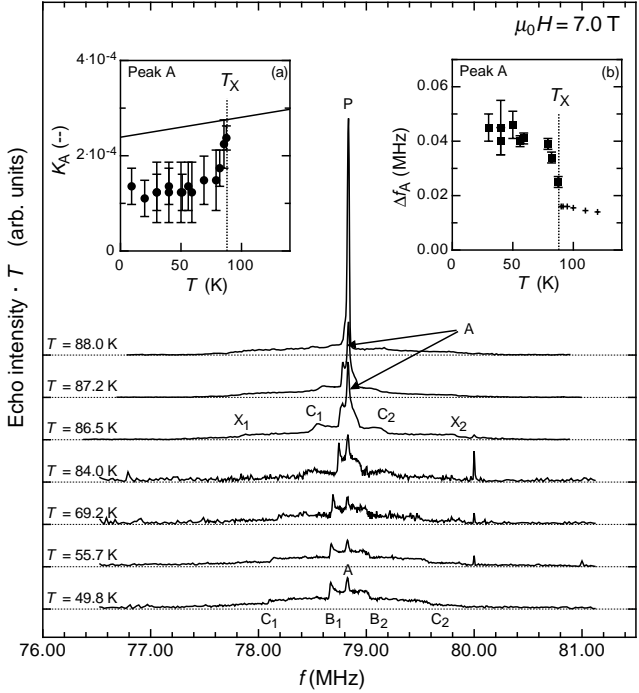}
\caption{
$^{23}$Na NMR spectra measured in a fixed magnetic field of $\mu_0H=7.0$ T, for $T\leq88.0$ K.
The integrated echo-intensity, multiplied with the temperature $T$, is plotted as a function of the irradiation frequency $f$.
The signifiance of the different letters is explained in the text.
Inset (a): Relative frequency shift $K_{\mathrm{A}}$ as a function of temperature $T$ of the $^{23}$Na NMR line A ($T\leq88$ K).
The solid line is the extrapolation for $T\leq88$ K of the solid line approximating $K_{\mathrm{P}}$ in Fig. \ref{FigKTParam}.
Inset (b): FWHM $\Delta f_{\mathrm{A}}$ as a function of temperature $T$ of the $^{23}$Na NMR line A ($T\leq88$ K). The small crosses represent $\Delta f_{\mathrm{P}}(T)$ ($T\geq88$ K).
}
\label{FigSpecAF}
\end{center}
\end{figure}

A generalized broadening of the NMR spectra, revealed by a substantial loss of intensity of the central line P, occurs at $T_{\mathrm{X}}$ upon cooling, as is shown in Fig. \ref{FigSpecAF} and in the inset of Fig. \ref{FigKTParam}.
At $T_{\mathrm{X}}$, peak P transforms into three different NMR signals, whose temperature evolution can be followed down to 30 K.
The three NMR signals are
\textit{(a)} a narrow signal labeled A,
\textit{(b)} a prominent rectangular signal in the range B$_1$-B$_2$, labeled B, and,
\textit{(c)} another rectangular signal, labeled C, in the range C$_1$-C$_2$.
Either the position or the ranges of the signals are indicated with the corresponding letters in the spectra displayed in Fig. \ref{FigSpecAF}.

The intensity of peak A is clearly smaller than that of the signals B and C.
The relative frequency shift $K_{\mathrm{A}}$ is represented in the inset (a) of Fig. \ref{FigSpecAF} as a function of temperature $T$.
Comparing $K_{\mathrm{A}}(T)$ with the extrapolation of $K_{\mathrm{P}}$ to low temperatures reveals a discontinuous change of slope at exactly $T_{\mathrm{X}}$.
The FWHM $\Delta f_{\mathrm{A}}$ of signal A exhibits an abrupt increase, followed by a trend to saturation to the constant value $\Delta f_{\mathrm{A}}\approx45$ kHz, which is reached at temperatures between 70 and 60 K (see inset (b) of Fig. \ref{FigSpecAF}).

\begin{figure}
 \begin{center}
  \leavevmode
  \epsfxsize=1.0\columnwidth \epsfbox{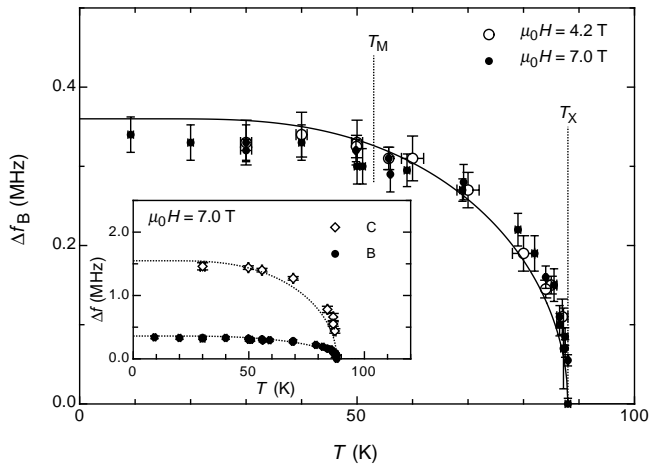}
\caption{
Width $\Delta f_{\mathrm{B}}$ of the rectangle-shaped signal B as a function of temperature $T$. $\Delta f_{\mathrm{B}}$ is extracted from the  $^{23}$Na NMR spectra measured in magnetic fields $\mu_0H=4.2$ T and $\mu_0H=7.0$ T.
The solid line represents the function (\ref{EqBrillouin}).
Inset: $\Delta f$ of the signals B and C.
}
\label{FigWidth}
\end{center}
\end{figure}

The effective frequency width $\Delta f_{\mathrm{B}}$ of signal B is calculated as
\begin{equation}
  \label{EqDeltaF}
  \Delta f_{\mathrm{B}} =f_{\mathrm{B}_2}-f_{\mathrm{B}_1}-\Delta f_{\mathrm{P}}^{\ast}
  \quad,
\end{equation}
where $f_{\mathrm{B}_1}$ and $f_{\mathrm{B}_2}$ are the frequencies of the lower and upper limits of the rectangular signal B,
and $\Delta f_{\mathrm{P}}^{\ast}=40$ kHz represents the total width in the paramagnetic state at $T_{\mathrm{X}}$.
In order to emphasize the influence of the ordering on the NMR signal, $\Delta f_{\mathrm{P}}^{\ast}$ has to be subtracted from the total width of signal B (see Eq. (\ref{EqDeltaF})).
In Fig. \ref{FigWidth} we present the temperature dependence of $\Delta f_{\mathrm{B}}$.
In the inset we compare $\Delta f_{\mathrm{B}}$ with $\Delta f_{\mathrm{C}}$, the latter being calculated in the same way as $\Delta f_{\mathrm{B}}$.
$\Delta f_{\mathrm{B}}$ and $\Delta f_{\mathrm{C}}$ both increase with decreasing temperature, approaching a constant value for $T\rightarrow0$.
From the main-frame diagram it is obvious that $\Delta f_{\mathrm{B}}$ is magnetic-field independent. The ratio
\begin{equation}
  \label{EqFreqRatio}
  \Delta f_{\mathrm{C}}/\Delta f_{\mathrm{B}}=4.2\pm 0.2
\end{equation}
is $H$- and $T$ -independent, indicating that the broadening of the two signals B and C is of common origin.
The behaviour of $\Delta f_{\mathrm{B}}(T)$ suggests that it can be considered to reflect an order parameter, as is indicated by the good agreement between the experimental data and
\begin{equation}
  \label{EqBrillouin}
  \Delta f_{\mathrm{B}}(T)=\Delta f_{\mathrm{B}}(0)\cdot \phi(T/T_{\mathrm{X}})
  \quad,
\end{equation}
with $\Delta f_{\mathrm{B}}(0)=0.36$ MHz, and $\phi(\alpha)$ representing the order-parameter type function, which is the solution of the functional equation $\phi(\alpha)=\tanh(\phi(\alpha)/\alpha)$.

\begin{figure}
 \begin{center}
  \leavevmode
  \epsfxsize=1.0\columnwidth \epsfbox{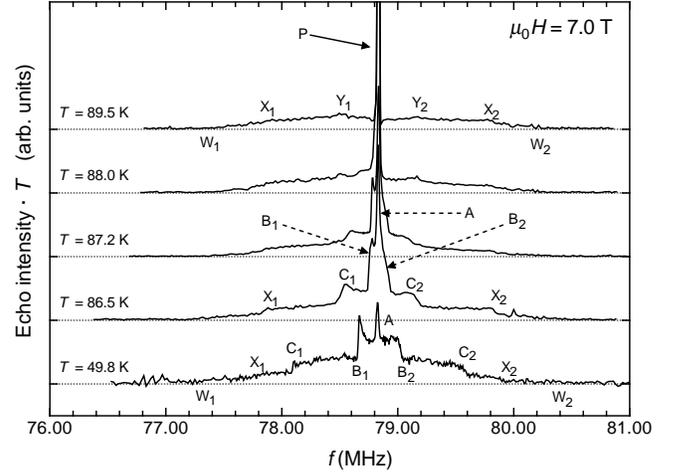}
\caption{
$^{23}$Na NMR spectra measured in a fixed magnetic field of $\mu_0H=7.0$ T, for $T$ between 86.5 and 89.5 K, and at $T=49.8$ K.
The integrated echo-intensity, multiplied with the temperature $T$, is plotted as a function of the irradiation frequency $f$.
The signifiance of the different letters is explained in the text.
}
\label{FigSpecTX}
\end{center}
\end{figure}

The variation of the spectra below $T_{\mathrm{X}}$ is thus attributed to the formation of a staggered internal field.
In particular, it is excluded that signals B and C are simply modifications of the quadrupolar wings $W$ observed at temperatures above $T_{\mathrm{X}}$.
Signals B and C emerge from the NMR line P, as is clearly recognized in Fig. \ref{FigSpecTX}.
The total intensity of the NMR signal (P+W above $T_{\mathrm{X}}$, A+B+C+W below $T_{\mathrm{X}}$) is unchanged within an error of 5$\%$ between 94.5 and 86.5 K, excluding a loss of signal intensity across the transition.
The wing features W$_1$, $X_1$ and W$_2$, $X_2$ are not affected by the phase transition at $T_{\mathrm{X}}$, yielding further support for the interpretation that signals B and C are due to transitions $+\frac{1}{2}\leftrightarrow -\frac{1}{2}$, and, in addition,  excluding the possibility of substantial charge rearrangements at $T_{\mathrm{X}}$.
The values of the staggered field at the B- and C-sites, respectively, are $\mu_0H_{\mathrm{int,B}}=\Delta f_{\mathrm{B}}(0)/2\gamma_{\mathrm{Na}}\approx0.016$ T and $\mu_0H_{\mathrm{int,C}}=4.1\times \mu_0H_{\mathrm{int,B}}\approx0.068$ T, respectively.

As may be seen in Figs. \ref{FigSpecAF} and \ref{FigSpecTX}, the quadrupolar features W$_1$, $X_1$ and W$_2$, $X_2$ persist with decreasing temperature down to 50 K.
The changes in their frequencies $f_{\mathrm{W}_1}$, $f_{\mathrm{W}_2}$, $f_{\mathrm{X}_1}$, and $f_{\mathrm{X}_2}$, as well as the rounding-off of their contours, are accomodated by the broadening of the signals B and C.

A peak-shaped anomaly in the spectra is observed at the low-frequency end of the rectangular signal B, i.e., at the feature labelled B$_1$ in Fig. \ref{FigSpecAF}.
We suspect that it is a deformation of the B-labeled rectangular signal, due to the partial alignment of grains of the sample in the magnetic field, or it may be an indication for a ferromagnetic component along the external field in the ordered state.

\subsection{Spin-lattice relaxation rate $T_1^{-1}$}

\begin{figure}
 \begin{center}
  \leavevmode
  \epsfxsize=1.0\columnwidth \epsfbox{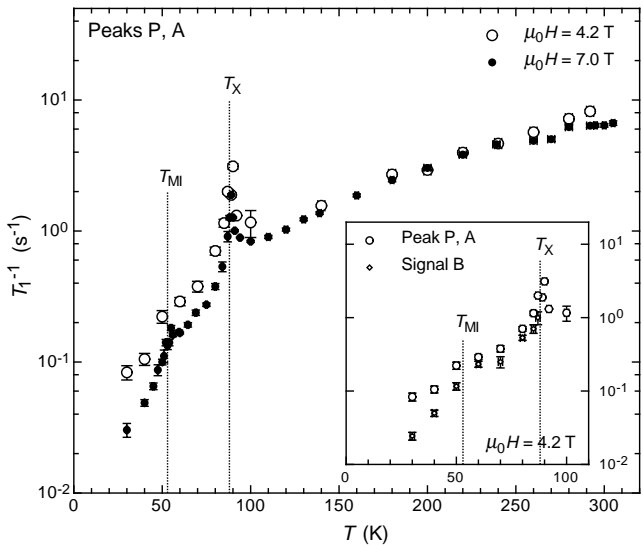}
\caption{
$^{23}$Na NMR spin-lattice relaxation rate $T_1^{-1}$ of the signals P ($T\geq88$ K) and A ($T\leq88$ K), as a function of temperature $T$, and in fixed magnetic fields $\mu_0H=4.2$ T and $\mu_0H=7.0$ T, respectively.
Inset: comparison of $T_1^{-1}(T)$ of the signals A and B, for temperatures below 100 K and in fixed magnetic field $\mu_0H=4.2$ T.
}
\label{FigT1}
\end{center}
\end{figure}

We measured the $^{23}$Na NMR spin-lattice relaxation rate $T_1^{-1}$ in the temperature range between 30 and 305 K and in fixed magnetic fields $\mu_0H=4.2$ T and $\mu_0H=7.0$ T.
$T_1^{-1}$ was obtained from monitoring the recovery of the $^{23}$Na nuclear magnetization $m$ after the application of a long comb of radiofrequency pulses. The experiments were performed by irradiating a frequency window of about 25 kHz. Above $T_{\mathrm{X}}$, the frequency was chosen to irradiate the NMR line P, while below $T_{\mathrm{X}}$ we centered the window such as to cover either line A or part of signal B. In all the cases, a double exponential recovery function \cite{Suter_1998_JPCM}
\begin{equation}
  \label{EqT1Recovery}
  m(t)=m_{\infty}\left[1-\left(0.4e^{-\frac{t}{T_1}}+0.6e^{-\frac{t}{6T_1}}\right)\right]
  \quad,
\end{equation}
characteristic of $I=3/2$ nuclei with only the central transition $+\frac{1}{2}\leftrightarrow -\frac{1}{2}$ irradiated, was observed.

In Fig. \ref{FigT1}, we display the temperature dependence of $T_1^{-1}$ and we note
a prominent peak in $T_1^{-1}(T)$ at $T_{\mathrm{X}}=88$ K.
Above $T_{\mathrm{X}}$, $T_1^{-1}(T)$ of the NMR line labeled P is field independent, as is expected in a state dominated by paramagnetic moments, but slowly decreases by a factor of 5 per 100 K upon cooling.
The fact that the magnetization recovery follows Eq. (\ref{EqT1Recovery}) supports the identification of signal P to be due to the central transition $+\frac{1}{2}\leftrightarrow -\frac{1}{2}$.

Below $T_{\mathrm{X}}$, $T_1^{-1}(T)$ exhibits an overall decrease by a factor of 20-50 per 100 K.
The comparison of the values obtained for $\mu_0H=4.2$ T and $\mu_0H=7.0$ T, respectively, indicates a decrease of $T_1^{-1}$ with increasing field at constant temperature, a dependence compatible with the scenario of magnetic ordering.
As shown in the inset of Fig. \ref{FigT1}, the ratio between the values of $T_1^{-1}$ of the signals A and B increases from 1 at 88 K to about 4 at 30 K.
This indicates that the two signals arise from different Na sites.
The general reduction of $T_1^{-1}(T)$ upon cooling is interrupted by a small cusp around the metal-insulator transition temperature $T_{\mathrm{MI}}=53$ K.

Several efforts were made to measure $T_1^{-1}(T)$ of signal C by irradiating a frequency window of about 25 kHz in the frequency ranges C$_1$-B$_1$ or B$_2$-C$_2$.
Because of the small magnitude of the recorded NMR signal, it was difficult to separate the contribution of the C-signal from perturbations arising from the quadrupolar wings of the B-signal, as well as from the $^{59}$Co NMR signal which, below $T_{\mathrm{X}}$, overlaps with the $^{23}$Na signal.
For this reason no reliable $T_1$ values could be obtained and these unsuccessful efforts were therefore abandoned. 

\subsection{Spin-spin relaxation rate $T_2^{-1}$}

\begin{figure}
 \begin{center}
  \leavevmode
  \epsfxsize=1.0\columnwidth \epsfbox{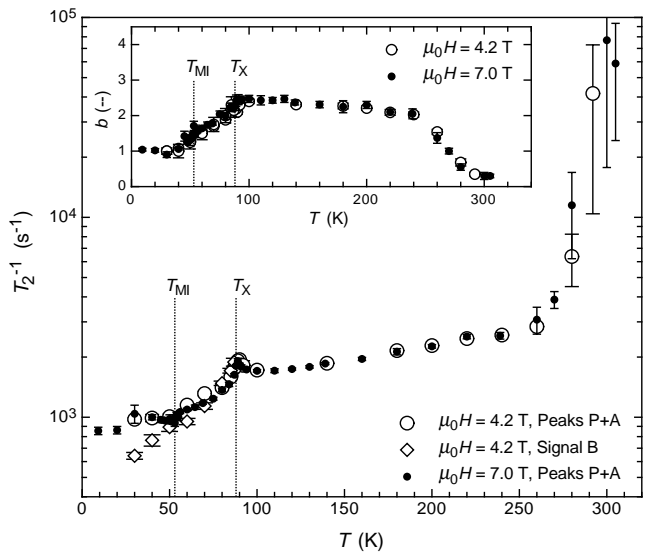}
\caption{
$^{23}$Na NMR spin-spin relaxation rate $T_2^{-1}$ as a function of temperature $T$ in fixed magnetic fields $\mu_0H=4.2$ T (signals P, A, and B) and $\mu_0H=7.0$ T (signals P and A only).
Inset: parameter $b$ appearing in the intensity decay (\ref{EqT2Recovery}), as a function of $T$.
}
\label{FigT2}
\end{center}
\end{figure}

Our measurements of the $^{23}$Na NMR spin-spin relaxation rate were made in the temperature range between 30 and 305 K and in fixed magnetic fields $\mu_0H=4.2$ T and $\mu_0H=7.0$ T.
The spin-echo lifetime $T_2^{\ast}$ was obtained by monitoring the spin-echo intensity as a function of the pulse delay time $\tau$ in the $\pi/2$-delay-$\pi$ echo sequence.
The irradiation conditions were chosen to be the same as in the $T_1$-experiments.
At all temperatures, the intensity $m(t)$ decayed according to the streched exponential behaviour
\begin{equation}
  \label{EqT2Recovery}
  m(t)=m(0)\exp\left(-(t/T^{\ast}_2)^{b}\right)
  \quad.
\end{equation}
Here, we used $t=2\tau+t_{\pi}+4/\pi\cdot t_{\pi/2}\approx 2\tau$, where $t_{\pi/2}$ and
$t_{\pi}$ are the durations of the $\pi/2$ and $\pi$ spin-echo pulses. 
$t$ corresponds to twice the time between the center of the $\pi$ pulse and the maximum of the echo signal
(see \cite{Slichter}, p. 45).
The spin-spin lattice relaxation rate $T_2^{-1}$ is approximated by
\begin{equation}
  T_2^{-1}\approx T_2^{\ast-1}
  \quad,
\end{equation}
which is justified because $T_2^{\ast-1}\gg T_1^{-1}$.

In Fig. \ref{FigT2} we display the temperature dependences of $T_2^{-1}$ and, in the inset, of the parameter $b$ appearing in the streched exponential function (\ref{EqT2Recovery}).
A value $b\neq1$ is attributed to indirect coupling between the nuclear spin at the different Na nuclei (for example, electron-mediated spin-spin interaction \cite{Ansermet_1988_JChemPhys,Walstedt_1995_PRB}), as opposed to direct scalar or dipolar interactions which yield $b=1$.
For example, a value $b=2$ was predicted and observed for the $^{63,65}$Cu NMR spin-echo decay in high-temperature superconductors \cite{Pennington_1989_PRB,Walstedt_1996_PRB}.
The exact value of $b$ depends on the details of the indirect coupling mechanism.
Nevertheless, a qualitative knowledge of the temperature dependence $b(T)$ is useful, because it can  reflect modifications in the interaction-mediating electron gas.

Four different temperature intervals can be identified, each characterized by its own $b(T)$ regime, and thus by its own electronic regime.
In the temperature range from 88 to 260 K, $T_2^{-1}$ is field independent and upon cooling slowly decreases by a factor of about 1.4 per 100 K.
The parameter $b\approx2.4$ is approximately constant.

Upon cooling, the peak in $T_2^{-1}(T)$ at $T_{\mathrm{X}}=88$ K is followed by an abrupt decrease.
Between 88 and 53 K, $T_2^{-1}$ is reduced by a factor of 6 per 100 K upon cooling, and the parameter $b$ decreases from $b\approx2.4$ to $b\approx1.3$.
The reduction of $T_2^{-1}$ when increasing the magnetic field from $\mu_0H=4.2$ T to $\mu_0H=7.0$ T is small (less than a factor of 1.1).
Differences between the values of $T_2^{-1}$ for the signals A and B are marginal (less than a factor 1.2).

Possible anomalies in $T_2^{-1}(T)$ at $T_{\mathrm{MI}}=53$ K are hidden by the noise in the data.
Below $T_{\mathrm{MI}}$, $T_2^{-1}(T)$ of signal A tends to a constant value with decreasing temperature, while for signal B, $T_2^{-1}(T)$ continues to decrease.
The parameter $b$ is approximatively unity below 40 K.

Above 260 K, an abrupt upturn in $T_2^{-1}(T)$ with increasing temperature is observed.
The enhancement is about a factor of 800 per 100 K.
The parameter $b$ changes from $b\approx2.4$ at 260 K to $b\approx0.5$ at 300 K.

\section{\label{SecResults}Discussion of the results}

\textit{Metallic phase above $T_{\mathrm{MI}}$.-}
We first turn our attention to the very unusual properties of Na$_{0.5}$CoO$_{2}$ in the metallic phase above $T_{\mathrm{MI}}$ and first discuss the magnetic susceptibility $\chi$.
At temperatures $T\geq T_{\mathrm{X}}$, $\chi(T)$ is more or less identical for Na$_{0.5}$CoO$_{2}$ and Na$_{0.31}$CoO$_{2}$ \cite{Foo_2004_PRL}, and is attributed to Pauli paramagnetism.
The decrease of $\chi(T)$ of about 20$\%$ between 300 and 100 K is thus not a singular feature of Na$_{0.5}$CoO$_{2}$.
The equivalence of the values of $\chi(T)$ above $T_{\mathrm{X}}$ for $x\leq0.5$ suggests that if the effective mass $m^{\ast}$ of the charge carriers is the same in all compounds, then their density is approximately the same across the range defined by $x\leq0.5$ of the Na$_{x}$CoO$_{2}$-series.
At this point, we note that the substantial enhancement of the effective electron mass for Na$_{0.5}$CoO$_{2}$, evaluated from the high-temperature saturation value of $\chi$ and cited above in Eq. (\ref{EqEffMass2D}), is an order of magnitude larger with respect to the result of an LSDA+U calculation \cite{Zhou_2005_PRL}.

A remaining puzzle of Na$_{0.5}$CoO$_{2}$ is the apparent inconsistency between magnetic and transport properties.
At room temperature, the conductivity $\sigma$ is of the same order of magnitude for Na$_{0.5}$CoO$_{2}$  and Na$_{0.31}$CoO$_{2}$ \cite{Foo_2004_PRL}, and, more generally, for all materials with $0.31\leq x\leq0.75$ .
This value implies a very short mean free path $l$, in particular for Na$_{0.5}$CoO$_{2}$, as cited above in Eq. (\ref{EqkFl2D}), and indicates an unusual transport regime.
It remains an open question whether a description of the transport properties in terms of a mean free path concept makes sense at all for Na$_{x}$CoO$_{2}$.
Although for $x=0.31$, and, in general, for $x$ different from 0.5, the conductivity $\sigma$ at least increases with decreasing temperature as expected for metals, for Na$_{0.5}$CoO$_{2}$, $\sigma(T)$ exhibits the opposite trend, indicating that magnetic and transport properties are not correlated in this case.

\textit{Evidence for magnetic ordering at $T_{\mathrm{X}}$.-}
The most instructive result of our experiments, the temperature evolution of the $^{23}$Na NMR spectra, provides a strong indication for Na$_{0.5}$CoO$_{2}$ to order antiferromagnetically at $T_{\mathrm{X}}=88$ K, as previously concluded on the basis of $\mu$SR-data \cite{Mendels_2005_PRL} and of results of Co NMR measurements \cite{Yokoi_2005_cond-mat}.

Above $T_{\mathrm{X}}$, the paramagnetic behaviour is reflected by the temperature- and field-independent proportionality $K_{\mathrm{P}}\sim \chi$.
Assuming an equal occupation of the sites Na1 and Na2, the appearance of a single NMR signal P suggests a uniform distribution of magnetic moment density in the conducting CoO$_2$-layers, implying a value of the hyperfine coupling constant $A_{\mathrm{P}}$ that is equal for all Na sites.

The broadening of signals B and C upon cooling below $T_{\mathrm{X}}$ (see Fig. \ref{FigWidth}) can only be accounted for by assuming the formation of a staggered internal field $H_{\mathrm{int}}$.
The magnetic nature of the transition is confirmed by the kink in $\chi(T)$ (see Fig. \ref{FigChi}) and by the anomalies in $T_1^{-1}(T)$ (see Fig. \ref{FigT1}), $T_2^{-1}(T)$ and $b(T)$ (see Fig. \ref{FigT2}) at $T_{\mathrm{X}}$.
The abrupt decrease of $T_1^{-1}(T)$ below $T_{\mathrm{X}}$ is typical for the suppression of magnetic fluctuations, which, in our case, is caused by the onset of an ordered state.
The magnetic ordering yields two inequivalent sites, B and C, with $H_{\mathrm{int,C}}/H_{\mathrm{int,B}}\approx4.2$.
From the ratio of the signal intensities B and C, which is close to 1, we argue that the two signals arise from two magnetically inequivalent but equally occupied Na sites.
Although other possibilities cannot be excluded a priori, it is plausible to identify the two sites with Na1 and Na2,  an assignment that results from previous work \cite{Huang_2004_PRB,Zandbergen_2004_PRB}.

\textit{Itinerant type of ordering at $T_{\mathrm{X}}$.-}
In comparison with the electronic-transport and magnetic-ordering features of a large number of oxide compounds \cite{Imada_1998_RMP}, Na$_{0.5}$CoO$_{2}$ seems to be a special case.
The magnetic order sets in at a temperature that significantly exceeds the metal-insulator transition temperature $T_{\mathrm{MI}}$, and not at $T_{\mathrm{MI}}$ \cite{Takano_1983} or below \cite{Lee_1997_PRL,Sternlieb_1996_PRL}.

From the analysis of the $\chi(T)$ data of Na$_{0.5}$CoO$_{2}$, presented in Sec. \ref{SecChi}, we conclude that at most a few percent of the Co ions carry a local magnetic moment which would be present also in the metallic phase.
This amount is certainly too small to identify the transition at $T_{\mathrm{X}}$ as an ordering of such moments upon cooling.
This implies that the observed antiferromagnetic order below $T_{\mathrm{X}}$ most likely involves the magnetic moments of the conduction electrons, and thus is of itinerant type.
Considering this situation, any attempt to determine the spontaneous magnetization density from NMR data alone is difficult and would yield only very speculative results.

\textit{Origin of signal A.-}
Despite its weak intensity, signal A is not attributed to the persistence of a small portion of paramagnetic phase in the sample.
This scenario would imply that signal A is a remnant of signal P.
The NMR data $K_{\mathrm{A}}(T)$, $\Delta f_{\mathrm{A}}(T)$ $T_{1,\mathrm{A}}^{-1}(T)$, and $T_{2,\mathrm{A}}^{-1}(T)$ below $T_{\mathrm{X}}$ (see Sec. \ref{SecNaNMR}) exclude this interpretation because their $T$-variations are correlated to those of signals B and C.
The latter are clearly features related to the ordered phase below $T_{\mathrm{X}}$.
Signal A may, e.g., be due to Na nuclei which couple more weakly to the ordered moments, because they are located at boundaries of domains with different magnetic and/or Na-ordering patterns.

\textit{Absence of charge disproportionation or structural modifications.-}
The temperature evolution of the $^{23}$Na NMR spectra around $T_{\mathrm{X}}$ ($T_{\mathrm{X}}\pm5$ K) provides no evidence for significant variations of the quadrupolar parameters $f_Q$ and $\eta$ (see Fig. \ref{FigSpecTX}).
This observation allows to straighten out a controversy, in the sense that we can exclude a charge disproportionation or a structural transition to occur at $T_{\mathrm{X}}$.

The features of the quadrupolar wings in the spectra can be tracked down to 50 K, i.e., to below the metal-insulator transition temperature $T_{\mathrm{MI}}$ (see Fig. \ref{FigSpecAF}).
We did not observe abrupt changes in these features at $T_{\mathrm{MI}}$.
This excludes an abrupt charge disproportionation with decreasing temperature at $T_{\mathrm{MI}}$.
The formal valence of the Co ions is either unaffected or, at most,  only gently altered by the metal insulator transition.

\textit{Magnetic component of the metal-insulator transition at $T_{\mathrm{MI}}$.-}
Our data reveal that the metal insulator transition also involves a magnetic component.
The abrupt decrease in $\chi(T)$ below $T_{\mathrm{MI}}=53$ K upon cooling, and the observed anomalies in $T_1^{-1}(T)$ and $T_2^{-1}(T)$ support this claim.
The metal-insulator transition is not reflected in the temperature evolution of the $^{23}$Na NMR spectra, indicating that the staggered magnetic field at the Na sites is not significantly modified at $T_{\mathrm{MI}}$.
A magnetic component of the metal-insulator transition was already claimed by the authors of Ref.\cite{Mendels_2005_PRL}.
Their conclusions were based on anomalous temperature dependences of two of the three detected $\mu$SR-frequencies at $T_{\mathrm{MI}}$, which were attributed to moderate reorientations of magnetic moments.
Such a scenario seems to be contradicted by our $T_1^{-1}$ data.
The saturation trend upon cooling exhibited by $T_1^{-1}(T)$ above $T_{\mathrm{MI}}$ is interrupted by an abrupt decrease in $T_1^{-1}(T)$ below $T_{\mathrm{MI}}$.
This behaviour cannot accomodated by small rearrangements of the magnetic moments, but more likely hints for the onset of a different regime of magnetic fluctuations below $T_{\mathrm{MI}}$.

\textit{Curie-Weiss behaviour of $\chi(T)$ at low temperatures.-}
In previous work \cite{Huang_2004_JPCM} an anomaly in the electrical conductivity $\sigma(T)$ at 25 K was observed and attributed to the occurrence of a phase transition.
Based on anomalies in the temperature evolution of $\mu$SR-frequencies at 29 K, the authors of Ref.\cite{Mendels_2005_PRL} suggested some rearrangements of magnetic moments in the ordered phase.
In view of these speculations, the observed upturn of $\chi(T)$ with decreasing temperature below 25 K (see Sec. \ref{SecChi}) requires some additional comments.
The presence of such an upturn was already reported in Ref.\cite{Foo_2004_PRL}, but other reports suggest that $\chi(T)$ tends to a constant value with $T\rightarrow0$ \cite{Huang_2004_JPCM,Yokoi_2005_cond-mat}.
The discrepancies suggest that the free magnetic moments, which determine the low-temperature Curie-Weiss behaviour of $\chi(T)$,  may be related to presence of defects or domains in the sample, whose concentration is sample dependent.
In any case, only a few percent of Co ions are involved.
Hence we argue that the upturn in $\chi(T)$ in Na$_{0.5}$CoO$_{2}$ is not related to any phase transition at all.

\textit{High-temperature regime.-}
Anomalies in $T_2^{-1}(T)$ and $b(T)$ are observed at 260 K (see Fig. \ref{FigT2}).
They are accompanied by the onset of a trend to saturation of $\chi(T)$ (see Fig. \ref{FigChi}).
However, we did not observe concomitant significant changes in the $^{23}$Na NMR spectra, implying that the cause of the above anomalies is very unlikely of magnetic type, but rather due to the onset of a different regime of interaction between the magnetic moments of the Na nuclei.
Since the onset of Na motion would be reflected by visible changes in the spectra, as was observed in Na$_{0.7}$CoO$_{2}$ \cite{Gavilano_2004_PRB,Gavilano_2005_Unpublished}, we attribute the anomalies in $T_2^{-1}(T)$ and $b(T)$ to small rearrangements in the Na positions.

\section{\label{SecConclusions}Conclusions}
In Na$_{0.5}$CoO$_{2}$, the onset of the insulating phase at $T_{\mathrm{MI}}=53$ K is preceded by an unusual type of antiferromagnetic ordering below $T_{\mathrm{X}}=88$ K, i.e., in the metallic phase.
Our data exclude a concomitant structural-type of ordering at $T_{\mathrm{X}}$.
The physical properties of the metallic phase above $T_{\mathrm{MI}}$ are characterized by a large effective mass and an extremely short mean free path of the itinerant charge carriers.
A convincing description of the itinerant-electron system of Na$_{0.5}$CoO$_{2}$ above $T_{\mathrm{MI}}$ is yet to be found.

\textit{Note added.-}
During the preparation of this manuscript, we got aware of a Co and Na NMR study on Na$_{0.5}$CoO$_{2}$ \cite{Bobroff_2005_cond-mat}.
The authors report clear evidence for antiferromagnetic ordering below 86 K, not accompanied by any charge disproportionation with decreasing temperature.
These conclusions are identical to those which follow from our experimental results.

\begin{acknowledgments}
We acknowledge useful discussions with M. Indergand and M. Sigrist.
We also acknowledge the support of K. Magishi in interpreting part of the NMR data.
The work benefitted from partial financial support of the Schweizerische Nationalfonds zur F\"{o}rderung der wissenschaftlichen Forschung. 
\end{acknowledgments}


\bibliography{Na05CoO2}

\end{document}